%% file: main.tex
\RequirePackage{amsmath}
\documentclass[twocolumn,conference,twoside]{IEEEtran}

\usepackage[utf8]{inputenc}
\usepackage{mathtools,amsfonts,amssymb,ragged2e}
\usepackage{pifont}
\usepackage{soul}
\usepackage[primitives,advantage,operators,sets,keys,adversary,probability,notions,events,complexity,landau,asymptotics]{cryptocode}
\usepackage{xspace}
\usepackage{hyperref}
\usepackage{cleveref}
\usepackage{xcolor}
\usepackage{mdframed}
\usepackage{acronym}
\usepackage{graphicx}
\usepackage{colortbl}
\usepackage{enumitem}
\usepackage{multirow}

\usepackage{array}
\usepackage{booktabs}
\usepackage{graphicx}
\usepackage{threeparttable}
\usepackage{wasysym}

\usepackage{epsfig}
\usepackage{calc}
\usepackage{amstext}
\usepackage{apalike}

\input{commands.tex}

\usepackage{fancyhdr}
\def\thesubsubsectiondis{\thesubsectiondis\arabic{subsubsection}}

\makeatletter
\def\subsubsection{%
  \@startsection
    {subsubsection}                 
    {3}                             
    {\parindent}                    
    {1.5ex plus 1.0ex minus 1.0ex}  
    {0.7ex plus .5ex minus 0ex}     
    {\normalfont\normalsize\itshape}
}
\makeatother

\begin{document}

\title{Authentication and Key Management Automation in Decentralized Secure Email and Messaging via Low-Entropy Secrets}

\author{\IEEEauthorblockN{Itzel Vazquez Sandoval\IEEEauthorrefmark{1},
 Arash Atashpendar\IEEEauthorrefmark{1}\IEEEauthorrefmark{2}, Gabriele Lenzini\IEEEauthorrefmark{1}}
\IEEEauthorblockA{\IEEEauthorrefmark{1}SnT, University of Luxembourg\\ \IEEEauthorrefmark{2}itrust consulting, Luxembourg\\
Email: \{itzel.vazquezsandoval,gabriele.lenzini\}@uni.lu, atashpendar.arash@gmail.com}}

\maketitle
\thispagestyle{fancy} 

\begin{abstract}
We revisit the problem of entity authentication in decentralized end-to-end encrypted email and secure messaging to propose a practical and self-sustaining cryptographic solution based on \ac{pake}. This not only allows users to authenticate each other via shared low-entropy secrets, e.g., memorable words, without a public key infrastructure or a trusted third party, but it also paves the way for automation and a series of cryptographic enhancements; improves security by minimizing the impact of human error and potentially improves usability.
First, we study a few vulnerabilities in voice-based out-of-band authentication, in particular a combinatorial attack against lazy users, which we analyze in the context of a secure email solution.
Next, we propose solving the problem of \emph{secure equality test} using \ac{pake} to achieve entity authentication and to establish a shared high-entropy secret key.
Our solution lends itself to offline settings, compatible with the inherently asynchronous nature of email and modern messaging systems. The suggested approach enables enhancements in key management such as automated key renewal and future key pair authentications, multi-device synchronization, secure secret storage and retrieval, and the possibility of post-quantum security as well as facilitating forward secrecy and deniability in a primarily symmetric-key setting. We also discuss the use of auditable \ac{pake}s for mitigating a class of online guess and abort attacks in authentication protocols.
\end{abstract}

\begin{IEEEkeywords}
Authentication, Key Management, Secure Email and Messaging, Password-Authenticated Key Exchange
\end{IEEEkeywords}

\section{\uppercase{Introduction}}
\label{sec:intro}
\input{intro}

\input{sota-and-structure}

\section{\uppercase{Framework and preliminaries}}
\label{sec:background}
\input{background}

\section{\uppercase{Pitfalls in out-of-band authentication}}
\label{sec:weakVoiceOOB}
\input{vulnerabilities}

\section{\uppercase{Authentication in email and messaging via PAKE}}
\label{sec:pake}
\input{pakes}

\section{\uppercase{Future work}}
\label{sec:conclusions}

\noindent
We foresee as a next step an implementation of our proposal, along with research on usability dedicated to assisting users with deriving low-entropy secrets, reducing mental effort and the likelihood of mistakes.
We also consider an analysis of our approach applied to encrypted mailing lists. Furthermore, we expect follow-up theoretical work on all the suggested cryptographic enhancements and implementations thereof.

\section*{ACKNOWLEDGEMENTS}
\noindent We thank Peter Y.A. Ryan for his valuable feedback and pEp Security S.A./SnT for funding the project “Security Protocols for Private Communications”.

\bibliographystyle{apalike}
\bibliography{pep-authentication}

\end{document}

%% file: commands.tex
\newcommand{\MA}{\mathcal{M}}
\newcommand{\adver}{\ensuremath{\MA}\xspace}
\newcommand{\A}{\ensuremath{\mathcal{A}}\xspace}
\newcommand{\B}{\ensuremath{\mathcal{B}}\xspace}

\newcommand{\G}{\ensuremath{\mathbb{G}}\xspace}

\renewcommand{\sk}{\ensuremath{\mathsf{sk}}\xspace}
\renewcommand{\pk}{\ensuremath{\mathsf{pk}}\xspace}
\newcommand{\Z}{\mathbb{Z}}
\newcommand{\kdf}[1]{\ensuremath{\mathsf{KDF}(#1)}}
\renewcommand{\mac}[1]{\ensuremath{\mathsf{MAC}(#1)}}

\newcommand{\sid}{\ensuremath{\mathsf{sid}}\xspace}
\newcommand{\txtmac}{\ensuremath{\text{MAC}}\xspace}

\newcommand{\pass}{\ensuremath{\pi}\xspace}
\newcommand{\tws}[1]{\mathsf{tws}(#1)}
\newcommand{\fpr}[1]{\mathsf{fpr}(#1)}

\newcommand{\cmark}{{\checkmark}}
\newcommand{\xmark}{{\ding{55}}}

\newcommand{\pep}{p$\equiv$p\xspace}

\acrodef{mitm}[MITM]{man-in-the-middle}
\acrodef{pep}[p$\equiv$p\xspace]{Pretty Easy Privacy}
\acrodef{otr}[OTR]{off-the-record messaging}
\acrodef{im}[IM]{instant messaging}
\acrodef{oob}[OOB]{out-of-band}
\acrodef{hisp}[HISP]{human interactive security protocol}
\acrodef{sas}[SAS]{short authentication string}
\acrodef{mac}[MAC]{message authentication code}
\acrodef{kdf}[KDF]{key derivation function}
\acrodef{dh}[DH]{Diffie-Hellman}
\acrodef{smp}[SMP]{socialist millionaires' problem}
\acrodef{ppss}[PPSS]{password-protected secret sharing}
\acrodef{pake}[PAKE]{password-authenticated key exchange}
\acrodef{mpc}[MPC]{multi-party computation}
\acrodef{pki}[PKI]{public key infrastructure}
\acrodef{ttp}[TTP]{trusted third party}
\acrodef{pq}[PQ]{Post-quantum}
\acrodef{ibe}[IBE]{identity-based encryption}
\acrodef{rom}[ROM]{random oracle model}
\acrodef{crs}[CRS]{common reference string}
\acrodef{icm}[ICM]{Ideal Cipher Model}
\acrodef{aam}[AAM]{Algebraic Adversary Model}
\acrodef{zk}[ZK]{zero-knowledge}
\acrodef{nizk}[NIZK]{non-interactive zero-knowledge}
\acrodef{cdh}[CDH]{Computational Diffie-Hellman}
\acrodef{csdh}[CSDH]{Computational Square Diffie-Hellman}
\acrodef{dsdh}[DSDH]{Decision Square Diffie-Hellman}
\acrodef{ddh}[DDH]{Decisional Diffie-Hellman}
\acrodef{didh}[DIDH]{Decision Inverted-additive Diffie-Hellman}
\acrodef{rlwe}[RLWE]{Ring Learning With Errors}
\acrodef{lwe}[LWE]{Learning With Errors}
\acrodef{svp}[SVP]{Shortest Vector Problem}
\acrodef{fs}[FS]{forward secrecy}
\acrodef{pfs}[PFS]{perfect forward secrecy}
\acrodef{kc}[KC]{key confirmation}

%% file: intro.tex
\noindent The use of email and \ac{im} has become pervasive and entrenched in the fabric of modern communication.
Thanks to cryptography, modern messaging tools have reached a considerable degree of sophistication (e.g., Signal) and offer advanced security features ranging from end-to-end encryption to forward secrecy and deniability. For these reasons, coupled with better usability, although email has a long history and remains undeniably popular with hundreds of billions of emails exchanged on a daily basis \cite{clark2018securing}, secure messaging has often been recommended by security experts as the go-to tool for secure communication.
Yet, secure messaging and email share two long-standing challenges, namely entity authentication and key management.

The primary concern is entity authentication, which invariably involves a mechanism that associates some cryptographic material with an identity, e.g., public key authentication. Key management, affecting email more acutely, is intertwined with authentication and the need for automating it has been known for a long time, e.g., see \cite{ruoti2018comparative}.

Over the years, several methods have been established to tackle authentication, and indirectly key management: manual validation, web of trust, \ac{pki} and hierarchical validation, public key directories as well as server-derived public keys such as \ac{ibe}. The set of viable candidates becomes much smaller once we consider a decentralized setting, i.e., without a \ac{pki} or a \ac{ttp}.
For this scenario, the body of work on key authentication contains hundreds of works focusing on methods based on the use of \ac{oob} channels and \ac{sas} comparisons, see \Cref{sec:related-work}. However, when it comes to schemes that rely on low-entropy shared secrets, which is what we address here, the only work that to the best of our knowledge proposes a solution is by \cite{alexander2007improved}. They use a modified solution to the \ac{smp} by \cite{boudot2001fair}, also known as secure equality test, for authentication in the \ac{otr} protocol.

Due to the required user interaction in most of these approaches, e.g., for verifying the authenticity of an interlocutor's public key, usability plays a key role in achieving authentication. Reducing the gap between security and usability, by finding optimal trade-offs, has been a central theme for decades with a plethora of long-standing open problems, e.g., see \cite{unger2015sokMessaging,clark2018securing}.

Here we revisit the problem of authenticating public keys in a decentralized setting and propose a user-friendly and robust approach based on \acf{pake} to solve \ac{smp} via low-entropy secrets. These secrets are not expected to be sampled from a large, uniformly distributed space, but rather from a small set of values, e.g., typical human-memorable passwords or pin numbers. The task of \ac{smp} boils down to two parties verifying equality of their inputs $\pass_\A$ and $\pass_\B$ in a zero-knowledge manner such that by the end they learn nothing but the boolean result of the test.
By solving \ac{smp} via \ac{pake}, we also establish a shared cryptographically strong secret key, making further cryptographic enhancements possible. Furthermore, this approach would not require any understanding of cryptographic concepts from the user, e.g., knowing about public-keys and fingerprints.

We also show how the suggested approach would not only work naturally in the context of secure messaging, but also in the inherently asynchronous setting of email. Apart from offering improved usability properties and eliminating a host of vulnerabilities present in \ac{oob}-based protocols, as discussed in \Cref{sec:weakVoiceOOB}, we show how the \ac{pake}-generated secret key can be used to pave the path towards providing a series of enhancements in secure email and messaging. These include inattentive user resistance, automated key renewal, automated future key pair authentication and multi-device synchronization, along with security properties such as deniability, forward secrecy, post-quantum security, auditability for detecting guess and abort attacks, secure secret storage and retrieval with applications in email and secure messaging.

By applying \ac{pake} to this problem, we advance the state-of-the-art in the use of shared low-entropy secrets for entity authentication, an idea considered only in \cite{alexander2007improved}. Also note that while \ac{smp} is a subproblem solved naturally by \ac{pake}, the latter has not been applied to the problem of authenticating public keys in decentralized settings.

\subsection{Motivation}
Entity authentication in decentralized, non-\ac{pki} environments is generally brushed aside. Solutions that do consider this problem typically rely on users correctly executing a manual comparison and even tend to keep this feature rather hidden, e.g., Signal.
Our incentive for replacing \ac{oob} authentication with a cryptographic protocol is the impact of failures occurring in methods highly-dependent on user behavior, which could completely jeopardize security.

Our motivation for using PAKE---a method that does not seem to have enjoyed enough recognition due to a lack of mature implementations, reluctance towards client side crypto, patent-encumbered designs and perhaps even unawareness of its usefulness---is grounded not only in its independence from a \ac{pki} or a \ac{ttp}, but also in its provision of a \ac{zk} solution for the secure equality test problem using a low number of rounds, compatible with asynchronous settings, and in the fact that it enables additional cryptographic enhancements.

\fancyhead[RO]{{\small Authentication and Key Management Automation in Decentralized Secure Email and Messaging via Low-Entropy Secrets}}
\fancyhead[LE]{Itzel Vazquez Sandoval, Arash Atashpendar and Gabriele Lenzini}

We were also motivated by two open problems stressed by
\cite{unger2015sokMessaging,clark2018securing}: bridging the gap between known theoretical results and real-world solutions, and the need for more robust authentication methods that also improve the trade-off between security and usability in secure solutions.
Finally, the need for addressing common challenges such as key management automation and device synchronization also spurred us on.

%% file: sota-and-structure.tex
\subsection{Contributions and structure}
After a brief review of the state-of-the-art in \Cref{sec:related-work}, we cover background concepts in \Cref{sec:background}.
In \Cref{sec:weakVoiceOOB}, we focus on a few vulnerabilities in the use of \ac{oob} channels for authentication, including a partial preimage attack aimed at lazy users, which we analyze in the context of the \pep secure email solution.
In \Cref{sec:pake}, we present an efficient \ac{pake}-based solution for authentication in secure messaging and email via low-entropy secrets, which enables further cryptographic enhancements.
We provide a concrete illustrative scheme along with an analysis of various \ac{pake} constructions and properties relevant for our work.
We show how our proposal can be used to achieve additional cryptographic tasks and properties such as automation in key management and key renewal, forward secrecy in a symmetric-key setting, deniability, post-quantum security, secure secret retrieval, and auditability for mitigating a certain class of online guess and abort attacks. We briefly analyze network transport mechanisms and security. \Cref{sec:conclusions} concludes with remarks on future work.

\subsection{Related Work}\label{sec:related-work}
The works of \cite{unger2015sokMessaging} and \cite{clark2018securing} provide extensive systematic surveys on secure messaging and email covering numerous aspects. We limit ourselves to the decentralized setting without elaborating on the drawbacks of web of trust approaches
covered in the above mentioned works.

The literature contains a sizeable body of work on \ac{oob}-based approaches, considered first by \cite{rivest1984expose}, many of which are inspired by the original work of \cite{vaudenay2005secure} based on \ac{sas} comparisons, e.g., \cite{nguyen2011authentication,kainda2009usability,kainda2010secureMobile,tan2017can}, to name a few.
This area has also been investigated by the formal methods community, see e.g. \cite{delaune2017formal} for a recent formal analysis of \ac{sas}-based schemes in the symbolic model.

As for low-entropy secret-based authentication, to the best of our knowledge, in the only work in the literature, \cite{alexander2007improved} use a modified version of a solution to \ac{smp} \cite{boudot2001fair},
which is mainly suitable for synchronous settings,
to improve authentication in \ac{otr} \cite{borisov2004off}.

%% file: background.tex
\noindent We use \A and \B to refer to honest parties Alice and Bob, and \adver for the adversary, Mallory. We use $\sample$ to denote an element sampled uniformly at random,
and $\concat$ to denote concatenation. We denote low-entropy secrets provided by users with \pass.

\textbf{Security model.}
We consider the standard Dolev-Yao model \cite{DolevYao}. We do not assume any trusted infrastructure. In one of our proposed methods for transport protocol, we assume the existence of untrusted buffer/relay servers, somewhat akin to the ones used in the design of Signal or OTR4 (see \Cref{sec:transport-mechanism}).
Regarding \ac{pake}s, we will consider various constructions in \Cref{sec:pake}, largely proven secure in the so-called BPR model \cite{bellare2000authenticated} under various hardness assumptions.

\textbf{Cryptographic notions.}
For space reasons, we assume familiarity with common cryptographic concepts, in particular with  \ac{dh}-based computational hardness assumptions.

We discuss schemes based on the \ac{rlwe} problem, a special case of the \ac{lwe} problem whose security may be reducible to the hardness of solving the \ac{svp} in lattices, for which no efficient quantum algorithms are known, thus conjectured to be quantum-secure. \ac{pq} cryptography encompasses schemes that are considered to be safe against adversaries equipped with scalable, cryptographically relevant quantum computers.

We use $\kdf{s}$ to denote a key derivation function that takes a source $s$ of keying material, typically with a fair amount of entropy but not uniformly distributed, and produces one or more cryptographically strong secret keys, see \cite{krawczyk2010cryptographic} for details. We denote with $\mac{k,m}$ a keyed message authentication code scheme that computes a tag on $m$ under key $k$.

\textbf{System requirements.}
We assume standard requirements for email transfer as our proposal does not require any format modifications and preserves compatibility between existing systems.
As for secure messaging, we do not introduce any extra trust assumptions and no additional infrastructure would be required. Any exchanges relayed or buffered by intermediate servers can be done by untrusted ones.

\textbf{Socialist Millionaires' Problem.}
In the realm of secure \ac{mpc}, Yao's millionaires' problem \cite{yao1982protocols} is a famous example in which two parties want to find out 
whose input is greater without revealing any more information on the actual value. \ac{smp} is a variant of this and a \ac{zk} proof of knowledge protocol, with the difference that the parties only wish to know if their inputs are equal.

A series of works have been dedicated to solving \ac{smp}, including a well-known solution by \cite{boudot2001fair} that provides a fair and efficient protocol, where fairness roughly means that no party can evaluate the function and walk away with the result without the other party learning the output.

\cite{garay2004efficient} showed that the fairness and the security definition of \cite{boudot2001fair} are not compatible with the simulation paradigm and that their solution would not be secure when composed concurrently; they present a construction that can be composed arbitrarily, with similar complexity results.

\textbf{PAKE.}
Password-authenticated key exchange (\ac{pake}) protocols enable the establishment of secure channels without the need for a \ac{pki}, \ac{ttp}s or empirical \ac{oob} channels. They allow two parties who share only a low-entropy secret, hereafter password, to agree on a cryptographically strong shared secret key, 
using the password for authentication.

Since the seminal work of \cite{DBLP:conf/sp/BellovinM92}, numerous \ac{pake} protocols have been proposed, which largely fall into the two categories of balanced (symmetric) and augmented (or asymmetric), referred to as a\ac{pake}. The latter stores one-way mappings of passwords on the server side in client-server settings.

Intuitively, a core property of \ac{pake} is that a run of the protocol should not leak any information about the password.
Moreover, they should be resistant to offline dictionary attacks; 
an online guessing attack with at most one test per run should be the optimal attack strategy for an active \adver interacting with a party. 
Similar to \ac{smp}, \adver can mask failed guessing attempts as network failures, thus allowing numerous attempts without raising suspicion. 
This is in general unavoidable, however, we will see in \Cref{sec:pake} how a recent work by \cite{roscoe2017auditable} can mitigate this.

%% file: vulnerabilities.tex
\noindent In \ac{oob} authentication, users compare some representation of a cryptographic
hash (fingerprint) of their partners' public keys via a separate authenticated channel.
This representation is usually in the form of a list of words, numbers or images.

Strong security and usability properties can be achieved if users execute the manual verification correctly.
Yet, the difficulty of having
users do the assigned tasks correctly while finding the right balance between usability and security is the root cause of security pitfalls, which have been amply discussed by research on fingerprint and \ac{sas} comparison via \ac{oob} channels (see \Cref{sec:related-work}).
Usability studies encourage the replacement of manual comparisons by automated software whenever possible \cite{tan2017can}.

\textbf{Selection of an adequate \ac{oob} channel.}
In practice, the theoretical and strong authentication requirements of \ac{oob} methods are not easy to satisfy. While face-to-face conversations provide a strong authenticated channel \cite{nguyen2011authentication}, they are often not viable.
It is usually assumed that an \ac{oob} channel cannot be forged, but it can be blocked, overheard, delayed or replayed.
Typical instantiations are done via voice-based channels, e.g. a phone call.
However, some already consider voice-based \ac{sas} comparison to be obsolete from a security perspective \cite{unger2015sokMessaging} as nowadays messages can be forged by voice synthesizers with a small sample of the victim's voice.
Indeed, a voice impersonation attack on users comparing PGP words \cite{wiretapping2014} reported the fake voice to be indistinguishable in about 50\% of the cases.

\textbf{Social engineering attacks.}
There are multiple ways for humans to interact via \ac{oob}, but with few indications about secure, privacy-preserving, or fair ways to do it, e.g., without knowing the authentication value, \adver can fool \A by pretending to be \B, asking her to read the words first, and confirming a full match.

\textbf{Inattentive and lazy users.} Users misreading words (inattentive) or comparing only subsets of them (lazy).
A recent paper by \cite{naor2018security}
analyzes approaches based on \ac{sas} authentication that are vulnerable to MITM attacks w.r.t. lazy users.
For instance, the approach in WhatsApp and Signal would be flawed if users compared only either the first or the second half of the value, since it would amount to verifying only one peer's fingerprint.
To fix this, the authors propose an influence spreading technique in which every bit of the value to be authenticated influences the generation of each element of the \ac{oob} representation.

\textbf{Partial preimage attack.}
\cite{dechand2016empirical} study an attack aimed at finding a partial preimage for a fingerprint verified by lazy users; specifically, they assume that subsets of bits at the boundaries and in the middle are checked.
Let $p$ denote the probability of finding a partial preimage for a given fingerprint $f$ and $q$ its complementary event.
To calculate $p=1-q$, we work out $q$ (i.e., the absence of partial preimages for a specific bit permutation).
Let $b$ be the length of the fingerprint $f$ and assuming that $r$ consecutive boundary bits are fixed (checked by the user), in this case, the leftmost and rightmost bits of $f$, we let $\ell$ denote the number of remaining bits in the middle from which a possible variation of $u$ bits could be fixed, i.e., checked by the user. Thus, we have $2 \cdot r + u$ fixed bits that the adversary cannot invert without the user noticing.
Valid preimages can thus be obtained by flipping up to $t = \ell - u$ bits within the middle bits; by removing these from the total space of size $2^b$, we obtain the number of invalid ones.
With $k$ denoting a given number of positions to modify, the valid strings are then given by $\binom{\ell}{k}$ choices of positions to flip. Thus, $q$ is given by
\begin{equation}
q = \frac{2^b-\sum_{k=1}^t\binom{\ell}{k}}{2^b}.
\end{equation}
Expressing $p$ as a function of the computational effort in terms of $e$ brute-force attempts, we have $p = 1 - q^e$. To estimate the number of steps needed for finding partial preimages with a success probability $\ge p$, we simply compute $e=\textrm{log}_q(1-p)$.
Expressing $e$ in base 2 gives results comparable to \cite{dechand2016empirical}.

\subsection{Case Study}\label{sec:sec-of-pep-auth}

Pretty Easy Privacy (\pep) is a software aimed at providing usable privacy-by-default in email via end-to-end opportunistic encryption.
The tool largely automates initial key generation and storage.
The public key of a user is attached to outgoing emails when a key of the recipient has not been stored.
Received keys are automatically stored for future use (\emph{trust-on-first-use}) and  outgoing emails are automatically encrypted when a public key of the intended receiver is available.
This approach requires neither a PKI nor a TTP.

Similar to the PGP word list, \emph{\pep trustwords} \cite{pepTwds} are natural language words that two users compare via a low-bandwidth \ac{oob} authenticated channel to prevent \ac{mitm} attacks.
The trustwords generation algorithm $\tws{\cdot}$ is a deterministic algorithm that runs locally taking as input the public key of the peer obtained by email and the user's own public key.
Informally, $\tws{\cdot}$ performs an $\textrm{XOR}$ over the fingerprints of each of the input arguments, and then maps each block of 16 bits from the resulting 160-bit long string to a word in a predefined dictionary of size $2^{16}$, thus yielding a list of ten words.

To encourage users to perform the \ac{oob} authentication, by default \pep shows only five words; this means that the peers compare the first 80 out of the 160 bits of a PGP fingerprint, assuming that they check all the words.
Since an ``influence spreading'' property, similar to Naor et al.'s, is already present, the best adversarial strategy is a brute-force attack over the public key space requiring $\bigO{2^{80}}$ steps to find a key $k$ such that the first 80 bits of $\fpr{k}$ are equal to those of $\fpr{\pk_B}$, with $\pk_B$ being the public key of \B.

We consider lazy users and compute estimates for partial preimage attacks similar to the one presented above.
We consider the two cases where, out of five words, the user verifies $(i)$ the first and last words as well as two from the middle $(ii)$ the first and last words, along with one of the three in the middle.
Let $b=80$, $\ell=48$ and for $(i)$ we have $u=32$ and we get $e \approx 2^{38}$; for $(ii)$, with $u=16$, we get $e \approx 2^{32}$. These results show that \adver would succeed with costs equal to and lower than the computational power estimated for an average adversary \cite{dechand2016empirical}.

Clearly the decision to show five words instead of ten by default needs to be reconsidered. Users might feel less annoyed by having to compare fewer words, however, its adverse effect on security is considerable as it practically renders brute-force attacks viable.

%% file: pakes.tex
\noindent We now show how \ac{pake} can be used to perform a secure equality test and thereby authentication.
Compared to \ac{otr} that uses a modified solution to the \ac{smp} protocol, we show that our \ac{pake}-based approach yields a more efficient solution with better security guarantees and enables further cryptographic features.

\textbf{Trust establishment using low-entropy secrets.}
For \A and \B to mutually authenticate, for now we assume that they share a low-entropy secret---e.g., a short password---either agreed upon beforehand or decided by posing and answering a question.
Intuitively, the goal is to perform a secure equality test such that upon termination of the protocol, \A and \B would only learn whether or not their respective secrets $\pi_A$ and $\pi_B$ were the same, thus authenticating each other on the basis of knowing the same secret.

In other words, \A and \B wish to authenticate their public keys via a secure equality test of their secrets without revealing any information about the latter, hence the need for a zero-knowledge protocol.
This means that the resulting transcript of their exchanges should not leak any information on $\pi_A$ and $\pi_B$ to \adver. Also, it should not be possible for \adver to brute-force the password via offline dictionary attacks. Thus, \adver's only strategy would amount to making online attempts.

\subsection{Public Key Authentication via PAKE}\label{sec:pake-based-auth}

To determine at the end of a \ac{pake} run whether the user secrets $\pi_A$ and $\pi_B$ are equal, without revealing anything else,
we suggest the enforcement of explicit authentication using \ac{kc} after the key establishment phase.
While this step may be optional in the general case for \ac{pake} protocols, here it would be necessary in order to bind the cryptographic material with an identity.
The information that \A and \B wish to authenticate---e.g., public keys for email addresses in \pep or key fingerprints for phone numbers in Signal---can be incorporated either into the \ac{kc} phase or into the initial user secrets.

Next we show using a concrete example how this can be constructed. For the moment, we do not focus on engineering aspects related to (a)synchronicity and message transport mechanisms, but we will come back to these in \Cref{sec:transport-mechanism}. The literature contains several well-studied instances of \ac{pake} and for this reason, we first pick a candidate to demonstrate how it can be used for public key authentication, and then compare a few prominent schemes according to specific properties of interest, as shown in \Cref{tab:pake}.

\subsubsection{An Instantiation based on SPAKE2}

\begin{figure*}[h]
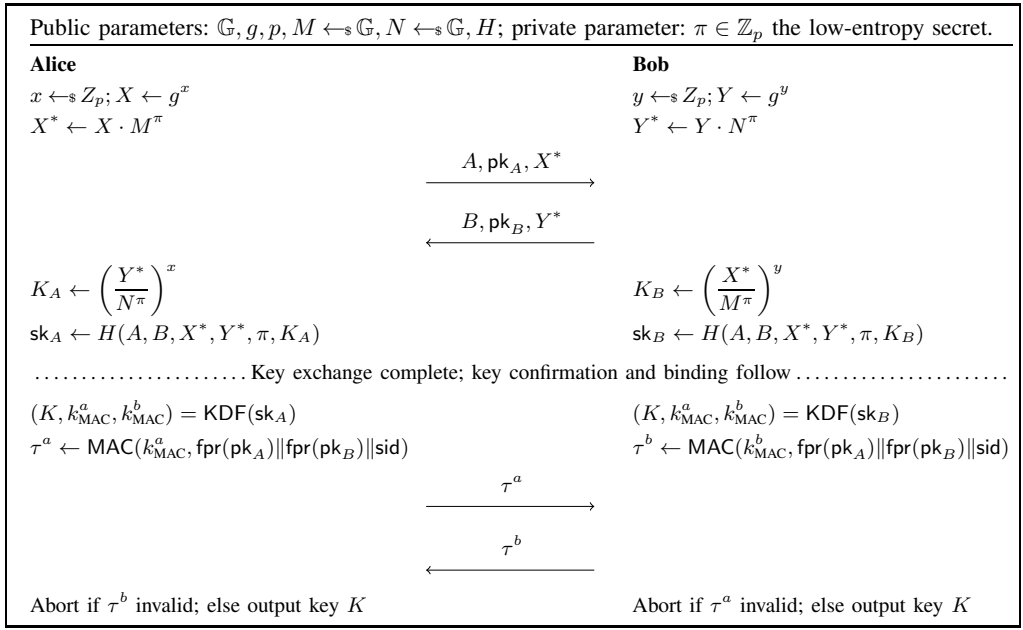

\centering
\resizebox{.75\textwidth}{!}{%
\fbox{
\procedure{Public parameters: $\G, g, p, M \sample \G, N \sample \G, H$; private parameter: $\pi \in \Z_p$ the low-entropy secret.}
{
\textbf{Alice} \> \> \textbf{Bob} \\
x \sample Z_p ; X \leftarrow g^x \> \> y \sample Z_p ; Y \leftarrow g^y\\
X^* \leftarrow X \cdot M^{\pi} \> \> Y^* \leftarrow Y \cdot N^{\pi} \\
\> \sendmessageright*[2.5cm]{A,\pk_A, X^*} \> \\
\> \sendmessageleft*[2.5cm]{B,\pk_B, Y^*} \> \\
K_A \leftarrow \left(\frac{Y^*}{N^{\pi}}\right)^x \> \> K_B \leftarrow \left(\frac{X^*}{M^{\pi}}\right)^y\\
\sk_A \leftarrow H(A,B,X^*,Y^*,\pi, K_A) \> \> \sk_B \leftarrow H(A,B,X^*,Y^*,\pi, K_B) \pclb
\pcintertext[dotted]{Key exchange complete; key confirmation and binding follow}
(K,k^a_{\txtmac},k^b_{\txtmac}) = \kdf{\sk_A} \> \> (K,k^a_{\txtmac},k^b_{\txtmac}) = \kdf{\sk_B} \\
\tau^a \leftarrow \mac{k^a_{\txtmac},\fpr{\pk_A} \concat \fpr{\pk_B} \concat \sid} \> \> \tau^b \leftarrow \mac{k^b_{\txtmac},\fpr{\pk_A} \concat \fpr{\pk_B} \concat \sid} \\
\> \sendmessageright*[2.5cm]{\tau^a} \> \\
\> \sendmessageleft*[2.5cm]{\tau^b} \> \\
\text{Abort if } \tau^b \text{ invalid; else output key } K \> \> \text{Abort if } \tau^a \text{ invalid; else output key } K
}}
}
\caption{\pk authentication using SPAKE2 with refresh-then-MAC key confirmation for entity binding.}
\label{fig:pep-spake2}
\end{figure*}

For illustration, in \Cref{fig:pep-spake2} we propose an extension of SPAKE2, a one-round protocol, with a \ac{kc} step to achieve explicit authentication, thus binding a public key to an entity. This yields a 2-round scheme, the minimum when \ac{kc} is enforced; see \cite{katz2011round} for optimal-round \ac{pake}s. For \ac{kc} we can use the generic refresh-then-\txtmac transformation.
Despite its long history and popularity, this transform
was only recently proved secure \cite{fischlin2016key}.

With $\G$ being a finite cyclic group of prime order $p$, generated by an element $g$, let $\G, g, p, M \sample \G, N \sample \G$ and hash function $H(\cdot)$ denote public parameters and $\pi \in \Z_p$ the private low-entropy secret, with the user password assumed to be appropriately mapped to an element in $\Z_p$.
The parties perform the key exchange phase, as shown in \Cref{fig:pep-spake2},
which concludes with the generation of a symmetric key. Upon termination of the key establishment, \A and \B each use the symmetric key to carry out a key-refreshing step via a key derivation function in order to generate fresh \txtmac keys (for both parties), along with a new session key, $K$, which will be the secret shared key.
Next, under the freshly generated keys, they each compute a \txtmac on the fingerprints of both parties' public keys.
The authentication now amounts to exchanging and verifying the obtained tags $\tau^a$ and $\tau^b$.

The addition of the \ac{kc} step increases the number of rounds and flows to 2 and 4, respectively. Note that this is merely an illustrative example and as already mentioned, other possibilities for \ac{kc} do exist, some of which offer additional properties. For instance, \cite{becerra2018forward} showed that a modified version of SPAKE2, called PFS-SPAKE2, coupled with a \ac{kc} step can achieve \ac{pfs} at the cost of increasing the number of rounds from 1 to 3. More recently, \cite{cryptoeprint:2019:1194abdalla} showed that SPAKE2 does indeed satisfy \ac{pfs} even without \ac{kc} under a different hardness assumption. They also prove a version with
a \ac{kc} step (yielding a better bound) almost identical to the one given in \Cref{fig:pep-spake2}, except that the protocol has one less flow.

Alternatively, the public key fingerprints can be embedded in the secret $\pass$, but note that even in that case, the \ac{kc} step cannot be skipped as an explicit authentication of the public keys would still be needed. More precisely, we would let
$\pass = \pass' \concat \fpr{\pk_\A} \concat \fpr{\pk_\B}$,
where $\pass'$ denotes the original user provided secrets, and we would compute the tags as $\tau^a \leftarrow \mac{k^a_{\txtmac},\sid}$, where the identifier \sid is computed over the transcript, with $\tau^b$ computed similarly.
Similar one round \ac{kc} methods for explicit authentication can be found in IETF internet-drafts for SPAKE2\footnote{\url{https://tools.ietf.org/id/draft-irtf-cfrg-spake2-08.html}} and J-PAKE\footnote{\url{https://tools.ietf.org/html/rfc8236}}.

\subsubsection{Choice of PAKE}\label{sec:pake-choice}

We consider a number of representative \ac{pake} protocols and analyze their properties w.r.t. our use case: SPAKE2 \cite{abdalla2005simple}, OPAQUE \cite{jarecki2018opaque}, PFS-SPAKE2 \cite{becerra2018forward}, J-PAKE \cite{hao2010j}, KV-SPOKE \cite{katz2011round}, RLWE-PAK and PPK \cite{ding2017provably}. \ac{pake}s are typically evaluated according to the security model in which they are proven secure, support for forward secrecy, the number of rounds, along with their communication and computational complexity. The complexity related aspects become more relevant in a client-server setting wherein a server has to process a high number of requests and sessions. In a decentralized peer-to-peer setting, such properties no longer play a major role.

In \Cref{tab:pake}, we present some of the relevant properties of the said constructions. Note that except for RLWE-PAK and RLWE-PPK that make use of lattice-based cryptography, all other schemes are Diffie-Hellman-based. In terms of \ac{pq} security, an immediate implication of this is that the latter cases would not be safe against quantum adversaries, whereas the first two would provide conjectured quantum-security due to the underlying \ac{rlwe} problem.

\begin{table}
\scriptsize
\centering
\begin{threeparttable}
\caption{Comparison of PAKE protocols.}
\label{tab:pake}
\begin{tabular}{l|c|c|c|c|c}
\hline
Protocol     & \multicolumn{1}{p{.7cm}|}{\centering Rounds/ \\ Flows}
             & KC
             & FS
             & \multicolumn{1}{p{.7cm}|}{\centering Security \\ model}
             & \multicolumn{1}{p{.7cm}}{\centering Hardness \\ assump.} \\
\hline
SPAKE2       & 1/2          & \xmark              & \cmark & ROM          & CDH    \\
PFS-SPAKE2   & 3/3          & \cmark              & \cmark & ROM          & CDH    \\
OPAQUE       & 2/3          & \cmark              & \cmark & ROM          & OMDH   \\
J-PAKE       & 2/4          & \xmark              & \cmark & ROM-AAM      & DSDH   \\
KV-SPOKE     & 1/2          & \xmark              & -      & CRS          & DDH    \\
RLWE-PAK     & 3/3          & \cmark              & \cmark & ROM          & RLWE     \\
RLWE-PPK     & 2/2          & \xmark              & \cmark & ROM          & RLWE     \\
\end{tabular}%
\begin{tablenotes}
\centering
\item ROM: Random Oracle Model;
    AAM: Algebraic Adversary Model;
    CRS: Common Reference String
\item DH: Diffie-Hellman;
        CDH: Computational DH;
        DDH: Decisional DH;
        DSDH: Decision Square DH;
        OMDH: One-More DH;
        RLWE: Ring Learning With Errors
\end{tablenotes}
\end{threeparttable}
\end{table}

Minimizing the number of rounds becomes more important for secure email than for messaging, especially if the transport mechanism is based on attachments or hidden emails (see \Cref{sec:transport-mechanism}).
As for secure messaging, this may be equally relevant for solutions that do not operate in a purely decentralized and peer-to-peer setting in which one may wish to reduce the load on relay or buffer servers, e.g., Signal or OTR4, but the number of rounds would in general be arguably less of a concern. Note that \ac{kc} can be added to schemes that do not have it by default at the cost of an extra round.

Intuitively, the notion of \ac{fs} captures the requirement that a long-term secret compromise should not result in prior session keys getting compromised and consequently the corresponding exchanges. Weak \ac{fs} (w\ac{fs}) refers to those schemes satisfying \ac{fs} against passive adversaries who did not interfere in the previous sessions and perfect \ac{fs} to those achieving the same against active adversaries. We will come back to this in \Cref{sec:crypto-properties}.

We limit the discussion on security models to practical considerations. In the \ac{rom}, an ideal truly random function being accessible to the parties through oracle calls is typically instantiated using cryptographic hash functions, and the \ac{crs} model implies the accessibility of a random string to all parties, generated in a trusted way. The latter may be less obvious to implement in the case of email due to the constraints of decentralization given that the generation of the \ac{crs} would be typically done by a trusted party or via a secure \ac{mpc} protocol, see e.g., \cite{sasson2014zerocash} for an example of \ac{crs} generation in a decentralized setting. Finally, regarding the RLWE-based schemes, their proofs are unfortunately in the \ac{rom}, as opposed to the quantum \ac{rom} (Q\ac{rom}), which would allow adversaries to query the random oracle in superposition.

\subsection{Cryptographic Enhancements}\label{sec:pake-enhancements}

We first show how a number of key properties related to key management automation and error resilience that have been identified in the literature \cite{unger2015sokMessaging} are satisfied and improved upon by our approach. We then present novel uses of \ac{pake}s in secure email and messaging. Note that once a \ac{pake}-generated key is established, subsequent \ac{pake} instances can be run automatically via a chaining self-sustaining mechanism.
While we mainly focus on enhancements for existing paradigms that depend on public-keys, we also consider possibilities for shifting to entirely symmetric-key solutions.
Indeed, once a \ac{pake}-generated shared symmetric key has been established, not only a wide range of well-understood techniques become possible, but one could also consider the benefits of transitioning to symmetric-key constructions, e.g., \txtmac-based authentication and symmetric-key encryption schemes.

In \Cref{tab:features}, we compare our proposal with a select set of approaches extracted from \cite{unger2015sokMessaging}. Due to space reasons, we refer the reader to the cited source for details on the properties therein.

\subsubsection{Key Management Automation}

\textbf{Automation of future key pair authentications.}
This is the underlying feature facilitating the achievement of some of the subsequent properties.
Once authentication between \A and \B is bootstrapped from an initial \ac{pake}, the authentication of new key pairs from either \A or \B can be automated using the \ac{pake}-generated shared keys without prompting the users to yet again enter new secrets.
For instance, in the case of email (e.g., \pep), authentication due to key pair generations can be triggered whenever a new key pair needs to be associated with an existing identity, or for binding a new email of \A or \B to a new key pair or when keys expire. Note that upon each future authentication, the \ac{pake}-generated symmetric keys can be refreshed by automatically carrying out a new \ac{pake}.

\textbf{Immediate enrolment.} This property refers to a user's keys being reinitialized in such a way that other parties can verify and use them immediately.
The \ac{pake}-generated key allows to automate the new key exchange and the corresponding authentication.

\textbf{Alert-less key renewal.}
Complementing the previous property, this one refers to users not receiving alerts or warnings prompting them to take action when other parties renew their public keys. This would be automated similarly to immediate enrolment.

\textbf{Low key maintenance.} This property pertains to the amount of user effort required for maintaining keys, e.g., tasks such as signing keys, renewing expired keys.
For instance, while the \pep client does automate key generation and renewal, the established trust level disappears with every key refreshment; key maintenance can be improved with PAKEs as explained above.

\textbf{Multi-device syncing.} Another quite natural application of \ac{pake} is in the realm of device pairing and multi-device synchronization. These typically rely on a \ac{hisp} and \ac{oob} techniques requiring manual intervention, which can give rise to new and subtle attacks. The application of \ac{pake}s in other contexts for device pairing has been considered before; it is thus natural to consider incorporating them in multi-device syncing of email agents and secure messaging systems.

\A can enter a password in both devices to be paired, $D_1$ and $D_2$, triggering a \ac{pake} protocol that establishes a secure channel between them for synchronization;
alternatively, this can even be done asynchronously without the two devices being online: $D_1$ pushes its state (e.g., key store, chat or email archive) to a server in encrypted form and later $D_2$ retrieves the secrets stored on the server in an oblivious manner w.r.t. the server, see \Cref{sec:secret-retrieval} for more details. For example, the current implementation of \pep resorts to an ad-hoc pairing technique for key synchronization that could benefit from such a \ac{pake}-based solution.

\textbf{Inattentive user resistance.} As discussed earlier, manual \ac{oob} key/fingerprint verification methods are susceptible to human error and inattentiveness. In the \ac{pake}-based approach, even if the users enter the wrong value, the result would not be as catastrophic as trusting a key prepared by the adversary.
At worst, it would be inconvenient as the authentication would fail prompting the user to eventually repeat the process.

\subsubsection{Cryptographic Properties}\label{sec:crypto-properties}

\textbf{Perfect forward secrecy (PFS).}
Once, more popular in the context of secure messaging (e.g., Signal and OTR), \ac{pfs} is now a requirement for cipher suites supported in TLS 1.3.
\ac{pfs} means that in the event of a password disclosure, prior derived session keys remain secure.
Clearly, a compromised \ac{pake}-generated key would have to be discarded and refreshed via a new \ac{pake} instance. A \ac{pake}-chaining mechanism that automatically performs key rotations and periodically refreshes the long-term key would provide limited windows of opportunity for \adver, after which the resulting key would be secure again.

Several \ac{pake} constructions provide \ac{pfs} by default, some of which are listed in \Cref{tab:pake}; it is known that \ac{pfs} can be obtained by adding explicit authentication via a \ac{kc} step \cite{bellare2000authenticated}.
This paradigm would be more relevant when such \ac{pake}-based approaches are used for synchronization purposes, be it device-to-device or device-to-server where \ac{pake} can be used to both
authenticate and establish a secure channel, thus providing \ac{pfs} for the session keys used for syncing. For more efficiency, a symmetric-key scheme with \ac{pfs} such as SAKE \cite{avoine2020symmetric} can be bootstrapped using \ac{pake}.

Finally, the approach adopted by the Sequoia-PGP project for adding \ac{fs} to OpenPGP-based solutions using regular sub-key rotations would also benefit from automated authentication in case the master key, certifying the short-term sub-keys, needs to be refreshed and authenticated. For additional security, with slightly hampered usability, a separation of storage can be enforced by for example storing such \ac{pake} long-term keys in dedicated hardware, e.g., hardware security modules or smart key storage devices such as YubiKey or Nitrokey, to protect against a device compromise; see \Cref{sec:secret-retrieval} for more details on this.

\textbf{Deniability.} This is another subtle and fundamental property that has been of particular interest in recent secure messaging systems such as Signal and OTR. Deniable exchange, applied to tasks ranging from authentication to encryption, has a long and somewhat controversial history due to the subtleties in various existing security definitions. We limit ourselves to the case of key exchange and the seminal framework of \cite{di2006deniable} providing security definitions in the simulation paradigm for deniable key exchange and authentication in which message and participation repudiation are considered as requirements.

Since limited space does not allow us to elaborate, we consider only sender/receiver deniability for non-augmented \ac{pake}, i.e., symmetric. We conjecture that such a construction would satisfy the said definition of deniability in the symmetric-key setting: in a two-party setup, a malicious party \adver would not be able to produce binding cryptographic proofs from communication transcripts, associating a party with a particular exchange, as all exchanges could have been simulated by the accusing party \adver. We now observe that a simulator can be constructed as $\pi$ is the only private input shared by both parties and all other parameters are public and drawn at random. Finally, assuming composability, using the \ac{pake}-generated key with symmetric ciphers and \txtmac-based authentication would preserve deniability. Clearly, this and other forms of deniability for \ac{pake} need to be studied rigorously in future work.

\textbf{Post-quantum security.} As pointed out in \Cref{sec:pake-choice}, in the event that secure messaging and email tools transition to \ac{pq} cryptography, there are candidate \ac{pake} constructions that provide conjectured \ac{pq} security; see \Cref{tab:pake}. Moreover, a \ac{pq}-secure \ac{pake} can be combined with the recent symmetric-key authenticated key exchange (SAKE) by \cite{avoine2020symmetric} that provides \ac{pfs} to obtain an efficient, \ac{pq}-secure and primarily symmetric key scheme with \ac{pfs}. A quantum-resistant \ac{pake} can be used once to bootstrap authentication via low-entropy secrets and to provide the initial master key needed by SAKE, which is conjectured to be \ac{pq}-secure due to its use of symmetric-key primitives, thus offering a low cost and efficient \ac{pq}-AKE suitable for settings with limited computational power.

\begin{table*}[t]
\newcommand*\rot[1]{\hbox to1em{\rotatebox[origin=bl]{65}{\textbf{#1}}} \hss}
\newcommand*\feature[1]{\ifcase#1 \xmark\or\LEFTcircle\or\CIRCLE\or\APLlog\or-\fi}
\newcommand*\f[3]{\feature#1&\feature#2&\feature#3}
\makeatletter
\newcommand*\ex[8]{#1&#2&%
    \f#3&\f#4&\f#5&\f#6&\f#7&\f#8&\expandafter\f\@firstofone
}
\makeatother
\newcolumntype{G}{c@{}c@{}c}
\caption{Comparison of trust establishment approaches.}
\label{tab:features}
\centering
\scriptsize
\begin{tabular}{lc G@{}G@{}G !{\kern0.5em} G@{}G@{}G !{\kern0.5em} G !{\kern.2em}}
\toprule
Paradigm  & Example & \multicolumn{9}{c}{Security} & \multicolumn{9}{c}{Usability} & \multicolumn{3}{c}{Adoption}\\
\midrule
&& \rot{Network MitM Prevented}
 & \rot{Operator MitM Prevented}
 & \rot{Operator MitM Detected}
 & \rot{Operator Accountability}
 & \rot{Key Revocation Possible}
 & \rot{Privacy Preserving}
 & \rot{Deniability Facilitated}
 & \rot{Forward Secrecy Facilitated}
 & \rot{Post-quantum Security}
 & \rot{Automatic Key Initialization}
 & \rot{Low Key Maintenance}
 & \rot{Easy Key Discovery}
 & \rot{Easy Key Recovery}
 & \rot{In-Band}
 & \rot{No Shared Secrets}
 & \rot{Alert-less Key Renewal}
 & \rot{Immediate Enrollment}
 & \rot{Inattentive User Resistant}
 & \rot{No Service Provider}
 & \rot{Asynchronous}
 & \rot{Multiple Key Support}\\
\midrule
\ex{Web of Trust}{PGP}   {222}{110}{040} {001}{100}{000} {222}\\
\ex{KD + SaL}{CONIKS}    {201}{222}{040} {222}{222}{222} {222}\\
\ex{OE + \ac{smp}}{OTR}  {111}{122}{440} {020}{011}{020} {202}\\
\ex{OE + TOFU}{TextSecure}  {111}{102}{444} {222}{222}{020} {220}\\
\ex{OE + TOFU + \ac{oob}}{\pep}  {111}{212}{444} {212}{102}{000} {200}\\
\cellcolor[gray]{.8}\ex{OE + TOFU + PAKE}{-}  {111}{212}{333} {222}{220}{222} {222}\\
\ex{KFV + \ac{oob}}{SilentText}  {222}{212}{444} {000}{002}{000} {000}\\
\cellcolor[gray]{.8}\ex{KFV + PAKE}{-}  {222}{212}{223} {222}{220}{222} {222}\\
\bottomrule
\end{tabular}

\smallskip
Paradigm property is: $\feature2=\text{satisfied}$;
     $\feature1=\text{partially satisfied}$;
     $\text{\feature0}=\text{not satisfied}$;
     $\feature3=\text{implementation dependent}$;
     $\text{\feature4}=\text{N/A}$

KD = Key directory; KFV = Key fingerprint verification; OE = Opportunistic encryption; SaL = Self-auditable logs; TOFU = Trust-on-first-use
\end{table*}

\subsubsection{Secure Secret Retrieval and Storage}\label{sec:secret-retrieval}

OPAQUE is a recent construction that can, among other things, serve as an a\ac{pake} to offer protection against breaches and server password file compromises.
It also offers a secret retrieval mechanism,  based on oblivious pseudo-random functions, to retrieve a secret from a server, stored in encrypted form, using only a low-entropy password.

This feature is inspired by the notion of \ac{ppss} schemes formalized by \cite{bagherzandi2011password}, which are $(t,n)$-threshold constructions wherein security is preserved against an adversary controlling up to $t$ servers out of $n$.
A problem that \ac{ppss} addresses is protecting \A's secret data $d$ (e.g., cryptographic secret key used for decryption, authentication credentials, etc) in the event of a device compromise.

Such a scheme would secret-share $d$ among a set of $n$ agents so that only a collusion of more than $t$ corrupt ones would compromise the data. Secret-sharing is combined with a password-based mechanism that allows the authentication of the owner of $d$ to the secret-share holders in order to trigger a reconstruction protocol and retrieve the secret.
The private storage of $d$ can be shared among $n$ external network entities to protect against user device compromise.
Alternatively, if \A does not trust external entities, her device can partake in the secret-sharing by storing multiple shares instead of any other external entity, thus preventing online dictionary attacks by a network attacker and not allowing \adver to learn anything about the secret without corrupting \A's device.

Secret retrieval would have several use cases in secure messaging.
For instance, instead of retrieving contacts from the user's phone, servers could store lists of contacts in encrypted form; this would enable asynchronous syncing of contacts across multiple devices without the service provider learning the content\footnote{It is worth noting that the developers of the Signal secure messaging protocol seem to have recently developed a similar functionality for the Signal application, which they refer to as ``Secure Value Recovery'' \cite{signalSVR2020}. Among other things, they describe a design involving a key stretching of a user's PIN and a master key derivation using the stretched key and a piece of server-side stored randomness. The same core functionality can be achieved using well-known solutions discussed in our work, i.e., the use of either \ac{ppss} or \ac{pake} constructions such as OPAQUE for secure secret storage and retrieval. Signal's developers also mention secret sharing and oblivious pseudo-random functions as future possibilities \cite{signalSVRtech2020}, both of which can be achieved using existing cryptographic primitives, as explained in this section.}.
A general anonymity/privacy related criticism directed at messaging services has to do with the identification of users via their phone numbers. This can be dealt with by securely storing long-term identities in encrypted form on the server, accessible only to the users.

Another use case would be to secret-share user data among several of their own devices, e.g., smartphone, laptop and tablet, so that a device compromise would not provide any useful information to an attacker; this can also be used for performing key synchronization among multiple devices. All these mechanisms would work in a similar manner from the user's point of view, i.e., simply by providing a password.

\subsubsection{Auditable \ac{pake}s for Thwarting Online Guessing Attacks}\label{sec:auditable-pake}

As is the case for \ac{smp} in \ac{otr}, online guessing attacks are unavoidable in \ac{pake}s. This is usually dealt with by fixing a limit on the number of failed attempts that can be tolerated before invalidating a password.

However, in certain cases, another subtle adversarial strategy aimed at sidestepping the (at most) one online test per run would be to resort to a class of guess and abort attacks in which \adver intercepts a message in a given session (or initiates a session of her own) at a crucial step of a protocol run, verifies her guess at the password and in case of an incorrect guess, drops the said message to disguise her attempt as a network communication failure.

This can be done in both directions to double the chance of discovering the password, or in parallel against many network nodes depending on the setting. Such an attack can be carried out repeatedly without raising an alarm as the honest parties may simply view this as a network failure.

We identify a similar vulnerability in the use of a modified version of \ac{smp} in \ac{otr}: just before the last phase where the parties perform their secure equality test, when \A and \adver exchange their blinded \ac{dh} terms incorporating the low-entropy password in the exponent, i.e., $(g_3^a,g_1^a g_2^{\pass_\A})$, \adver could make a guessing attempt at $\pass_\A$ and in case of obtaining 0 (not equal), drop the message and force an abort, see sections 4.2 and 4.3 in \cite{alexander2007improved}. Note that the \ac{nizk} proofs that are attached to the messages at every exchange are not meant to protect against this type of attack.

In a relatively recent work, \cite{roscoe2017auditable} apply a mechanism based on commitment schemes and delay functions (e.g., timed-release encryption), originally developed by \cite{roscoe2016detecting} for protecting against online attacks in \ac{hisp}s that use \ac{sas}, to the setting of \ac{pake}s in order to make them auditable by achieving \emph{stochastic fair exchange}.

Roughly speaking, this is achieved by a transformation for \ac{pake}s at the level of \ac{kc} using a combination of blinding, randomization, commitments and delay functions such that a series of messages consisting of fake ones and the real intended message are exchanged and the parties will only get to know which is the \emph{right} one until their exchange is complete. In a follow-up work, \cite{cryptoeprint:2019:1281} generalize this result to achieve $\varepsilon$-fair exchange using oblivious transfer and timed-release encryption.

This transformation can be used to enhance any \ac{pake} with auditability, thus lending itself quite naturally to the authentication method suggested in this work.
An important limitation here is that, due to the highly interactive design of the solution, it would be more suitable to the setting of secure messaging than email, unless a given email solution were to opt for untrusted buffer servers for transport, see \Cref{sec:transport-mechanism}.

Finally, note that some of the ideas in this transformation, specifically those related to enforcing fairness, have common elements with the original \ac{smp} \cite{boudot2001fair} solution aimed at providing fairness, a property that was removed from the modified version of \ac{smp} used in \ac{otr} \cite{alexander2007improved} on account of achieving efficiency.

\subsection{Transport Mechanism}\label{sec:transport-mechanism}

\textbf{Email-based approach.} Given the small number of rounds required by PAKE protocols, in the case of email we can afford to use standard email attachments or specially formatted hidden emails as messages' carriers, processed by the email client in the background.

\A can choose (via an interface option) to enter her secret $\pass_\A$ upon sending her first email, allowing the first flow of the protocol to occur via an attachment; similarly, when \B replies, if he opts for entering his secret $\pass_\B$, the initial \ac{pake} round would be done; the subsequent \ac{kc} can be done automatically by the dedicated software.

Alternatively, one can resort to a hidden email transport model such as the one used by \pep for multi-device key synchronization. Here, the implementations would encapsulate crypto messages in specially crafted email attachments, kept hidden from the user (e.g., archived separately) and processed automatically. Since we primarily deal with authentication, our proposal would have minimal impact in terms of communication and computational complexity as it would have to take place only once.

\textbf{Untrusted server approach.}
Although early \ac{im} tools were entirely online services that maintained an active session for each conversation, modern \ac{im} tools are in fact quite similar to email in that the underlying system follows an asynchronous model.
Both Signal and the latest version of \ac{otr} \cite{otrv4Spec} achieve offline messaging by using ``buffer servers'' for hosting pre-key bundles that can be fetched without the other party being online.

We can use a similar mechanism to overcome transport engineering obstacles in email more elegantly,
since all aspects related to the exchange of emails remain unchanged and thus interoperable. In fact, the use of an intermediate server would not introduce additional trust assumptions as the transcript of a PAKE protocol does not leak useful information to the adversary; such a server would be untrusted and any entity would be able to set up their own instance.

\subsection{Security and Low-Entropy Secrets}

The schemes considered thus far come with proofs of security, see \Cref{tab:pake} for the corresponding models and assumptions. The security guarantees can be traced back to the core properties of \ac{pake}s: they can in effect fulfill the role of \ac{zk} proof of knowledge schemes such that a run of the protocol does not leak any information on the password and upon termination only reveals whether the secrets were equal; they resist offline dictionary attacks and online ones by limiting active adversarial tests to one password per run; compromised session keys will not compromise the security of other session keys; depending on the choice of \ac{pake}; \ac{fs} would ensure that past session keys remain secure if the password is leaked.

The only way for \adver to gain knowledge about the secret would be via active online guessing attempts, typically dealt with by fixing a limit on the number of failed attempts, e.g., \ac{smp} in \ac{otr}. We discussed how the possibility of making \ac{pake}s auditable can be used to mitigate this class of attacks by distinguishing between failed adversarial attempts and network failures to minimize the adversary's tries to one, under the assumption of correct input entry by honest users.

\textbf{Low-entropy secret agreement.}\label{sec:inband-secret}
Our proposal does come with its own caveat, namely the need for either presharing or agreeing on a low-entropy secret in-band. As already discussed in \cite{alexander2007improved}, the users can either share a secret over a secure channel, e.g. \ac{oob}, or agree on one via an in-band solution without revealing sensitive information about the secret itself, for instance, \A asking \B to use the name of their favorite restaurant. The user interface of a tool implementing this could warn users not to include the secret itself, similar to standard email warnings reminding users to attach documents in case they have mentioned it in the body of the message.

Another possibility would be to use another already authenticated and secure channel to agree on a secret.
For instance, given the widespread use of tools such as Signal, the parties could simply use it to agree on a secret for a one-time entity authentication of their secure email solution. While it may not be appealing from a theoretical point of view, due to the assumption of there being an already authenticated and secure channel, practically speaking, this approach would in fact provide a realistic and usable solution.

\textbf{Usability aspects.} Implementations of the approach must pay proper attention in providing an adequate interface for entering the low-entropy secret, in addition to the usual considerations for providing easy explanations and documentation for users.
A lesson learned from a usability study on the \ac{otr}/\ac{smp} tool \cite{usabilityOTR} stresses the need for further research on how to guide users towards establishing a secure shared human-memorable secret.

For instance, adding a list pre-populated with questions might serve as a guide to generate similar ones or reduce user effort by allowing them to choose one from the list; the questions should not lead to evident answers or to answers belonging to very small known sets, such as ``yes/no'' or colors, as such cases increase the successful guessing probability of the adversary. Another measure for dealing with disparities due to letter cases would be to just convert the secret to upper-case, at the cost of reducing entropy.